\documentclass[]{aastex701}

\usepackage{amssymb}
\usepackage{array}
\usepackage{booktabs}

\setcitestyle{numbers,square}

\makeatletter
\renewcommand{\frontmatter@title@above}{%
  \footnotesize
  \textsc{\@journalinfo}\par
}
\makeatother
\journalinfo{Rencontres de Moriond, Cosmology, 2026 \vspace{0.5cm}}

\begin{document}
\correspondingauthor{C.~Verg\`{e}s}

\title{Controlling instrumental systematics for the BICEP inflation survey}

\author[0000-0002-3942-1609]{C.~Verg\`{e}s}%
\affiliation{Physics Division, Lawrence Berkeley National Laboratory, Berkeley, CA 94720, USA}%
\email[show]{cverges@lbl.gov}%
\author{P.~A.~R.~Ade}%
\affiliation{School of Physics and Astronomy, Cardiff University, Cardiff, CF24 3AA, United Kingdom}%
\email{Peter.Ade@astro.cf.ac.uk}%
\author[0000-0002-9957-448X]{Z.~Ahmed}%
\affiliation{Kavli Institute for Particle Astrophysics and Cosmology, Stanford University, Stanford, CA 94305, USA}%
\affiliation{SLAC National Accelerator Laboratory, Menlo Park, CA 94025, USA}%
\email{zeesh@slac.stanford.edu}%
\author[0000-0001-6523-9029]{M.~Amiri}%
\affiliation{Department of Physics and Astronomy, University of British Columbia, Vancouver, British Columbia, V6T 1Z1, Canada}%
\email{mandana@physics.ubc.ca}%
\author[0000-0002-8971-1954]{D.~Barkats}%
\affiliation{Center for Astrophysics, Harvard \& Smithsonian, Cambridge, MA 01238, USA}%
\email{dbarkats@cfa.harvard.edu}%
\author[0000-0002-3351-3078]{R.~Basu~Thakur}%
\affiliation{Department of Physics, California Institute of Technology, Pasadena, CA 91125, USA}%
\email{ritoban@caltech.edu}%
\author[0000-0001-9185-6514]{C.~A.~Bischoff}%
\affiliation{Department of Physics, University of Cincinnati, Cincinnati, OH 45221, USA}%
\email{bischocn@ucmail.uc.edu}%
\author[0000-0003-0848-2756]{D.~Beck}%
\affiliation{Department of Physics, Stanford University, Stanford, CA 94305, USA}%
\email{dobeck@stanford.edu}%
\author{J.~J.~Bock}%
\affiliation{Department of Physics, California Institute of Technology, Pasadena, CA 91125, USA}%
\affiliation{Jet Propulsion Laboratory, California Institute of Technology, Pasadena, CA 91109, USA}%
\email{jjb@astro.caltech.edu}%
\author{H.~Boenish}%
\affiliation{Center for Astrophysics, Harvard \& Smithsonian, Cambridge, MA 01238, USA}%
\email{hans.boenish@gmail.com}%
\author{V.~Buza}%
\affiliation{Kavli Institute for Cosmological Physics, University of Chicago, Chicago, IL 60637, USA}%
\email{vbuza@kicp.uchicago.edu}%
\author[0000-0003-4541-7080]{B.~Cantrall}%
\affiliation{Department of Physics, Stanford University, Stanford, CA 94305, USA}%
\affiliation{Kavli Institute for Particle Astrophysics and Cosmology, Stanford University, Stanford, CA 94305, USA}%
\email{cantrall@stanford.edu}%
\author[0000-0002-1630-7854]{J.~R.~Cheshire~IV}%
\affiliation{Department of Physics, California Institute of Technology, Pasadena, CA 91125, USA}%
\email{cheshire@caltech.edu}%
\author{J.~Connors}%
\affiliation{National Institute of Standards and Technology, Boulder, CO 80305, USA}%
\email{jake.connors@nist.gov}%
\author[0000-0002-2088-7345]{J.~Cornelison}%
\affiliation{High-Energy Physics Division, Argonne National Laboratory, Lemont, IL, 60439, USA}%
\email{james.a.cornelison@gmail.com}%
\author{M.~Crumrine}%
\affiliation{School of Physics and Astronomy, University of Minnesota, Minneapolis, MN 55455, USA}%
\email{crumrine@umn.edu}%
\author{A.~J.~Cukierman}%
\affiliation{Department of Physics, California Institute of Technology, Pasadena, CA 91125, USA}%
\email{ajcukier@caltech.edu}%
\author{E.~Denison}%
\affiliation{National Institute of Standards and Technology, Boulder, CO 80305, USA}%
\email{ed.denison@nist.gov}%
\author{L.~Duband}%
\affiliation{Service des Basses Temp\'eratures, Commissariat \`a l'\'Energie Atomique, 38054 Grenoble, France}%
\email{lionel.duband@cea.fr}%
\author[0000-0002-7059-8728]{M.~A.~Echter}%
\affiliation{Center for Astrophysics, Harvard \& Smithsonian, Cambridge, MA 01238, USA}%
\email{mechter@cfa.harvard.edu}%
\author[0009-0007-6718-1730]{M.~Eiben}%
\affiliation{Faculty of Physical Sciences, University of Iceland, 102 Reykjavík, Iceland}%
\email{mirandaeiben@gmail.com}%
\author[0000-0003-4117-6822]{B.~D.~Elwood}%
\affiliation{Center for Astrophysics, Harvard \& Smithsonian, Cambridge, MA 01238, USA}%
\affiliation{Department of Physics, Harvard University, Cambridge, MA 02138, USA}%
\email{bdelwood@fas.harvard.edu}%
\author[0000-0002-3790-7314]{S.~Fatigoni}%
\affiliation{Department of Physics, California Institute of Technology, Pasadena, CA 91125, USA}%
\email{sofiaf@caltech.edu}%
\author[0000-0001-8217-6832]{J.~P.~Filippini}%
\affiliation{Department of Physics, University of Illinois at Urbana-Champaign, Urbana, IL 61801, USA}%
\email{jpf@illinois.edu}%
\author{A.~Fortes}%
\affiliation{Department of Physics, Stanford University, Stanford, CA 94305, USA}%
\email{afortes@stanford.edu}%
\author{M.~Gao}%
\affiliation{Department of Physics, California Institute of Technology, Pasadena, CA 91125, USA}%
\email{mingao@caltech.edu}%
\author{C.~Giannakopoulos}%
\affiliation{Department of Physics, University of Cincinnati, Cincinnati, OH 45221, USA}%
\email{giannacs@ucmail.uc.edu}%
\author{N.~Goeckner-Wald}%
\affiliation{Department of Physics, Stanford University, Stanford, CA 94305, USA}%
\email{ngoecknerwald@gmail.com}%
\author[0000-0001-5268-8423]{D.~C.~Goldfinger}%
\affiliation{Department of Physics, Stanford University, Stanford, CA 94305, USA}%
\email{dgoldfin@stanford.edu}%
\author{S.~Gratton}%
\affiliation{Centre for Theoretical Cosmology, DAMTP, University of Cambridge, Cambridge CB3 0WA, UK}%
\affiliation{Kavli Institute for Cosmology Cambridge, Cambridge CB3 0HA, UK}%
\email{stg20@cam.ac.uk}%
\author{J.~A.~Grayson}%
\affiliation{Department of Physics, Stanford University, Stanford, CA 94305, USA}%
\email{jmsgrysn@gmail.com}%
\author[0009-0003-6999-0129]{A.~Greathouse}%
\affiliation{Department of Physics, California Institute of Technology, Pasadena, CA 91125, USA}%
\email{anns@caltech.edu}%
\author[0000-0001-9292-6297]{P.~K.~Grimes}%
\affiliation{Center for Astrophysics, Harvard \& Smithsonian, Cambridge, MA 01238, USA}%
\email{pgrimes@cfa.harvard.edu}%
\author[0000-0003-2221-3018]{G.~Halal}%
\affiliation{Department of Physics, Stanford University, Stanford, CA 94305, USA}%
\email{georgech@stanford.edu}%
\author{M.~Halpern}%
\affiliation{Department of Physics and Astronomy, University of British Columbia, Vancouver, British Columbia, V6T 1Z1, Canada}%
\email{halpern@physics.ubc.ca}%
\author{E.~Hand}%
\affiliation{Department of Physics, University of Cincinnati, Cincinnati, OH 45221, USA}%
\email{handem@mail.uc.edu}%
\author{S.~A.~Harrison}%
\affiliation{Center for Astrophysics, Harvard \& Smithsonian, Cambridge, MA 01238, USA}%
\email{samuel.a.harrison@gmail.com}%
\author{S.~Henderson}%
\affiliation{Kavli Institute for Particle Astrophysics and Cosmology, Stanford University, Stanford, CA 94305, USA}%
\affiliation{SLAC National Accelerator Laboratory, Menlo Park, CA 94025, USA}%
\email{shawn@slac.stanford.edu}%
\author[0000-0002-3437-5228]{T.~D.~Hoang}%
\affiliation{School of Physics and Astronomy, University of Minnesota, Minneapolis, MN 55455, USA}%
\email{hoang416@umn.edu}%
\author{J.~Hubmayr}%
\affiliation{National Institute of Standards and Technology, Boulder, CO 80305, USA}%
\email{johanneshubmayr@gmail.com}%
\author[0000-0001-5812-1903]{H.~Hui}%
\affiliation{Department of Physics, California Institute of Technology, Pasadena, CA 91125, USA}%
\email{hhui@caltech.edu}%
\author{K.~D.~Irwin}%
\affiliation{Department of Physics, Stanford University, Stanford, CA 94305, USA}%
\email{irwin@stanford.edu}%
\author{M.~Izquierdo~Poza}%
\affiliation{Department of Physics, Boston University, Boston, MA 02215, USA}%
\email{marcelai@bu.edu}%
\author[0000-0002-3470-2954]{J.~H.~Kang}%
\affiliation{Department of Physics, California Institute of Technology, Pasadena, CA 91125, USA}%
\email{jkang7@caltech.edu}%
\author[0000-0002-5215-6993]{K.~S.~Karkare}%
\affiliation{Department of Physics, Boston University, Boston, MA 02215, USA}%
\email{kkarkare@bu.edu}%
\author{S.~Kefeli}%
\affiliation{Department of Physics, California Institute of Technology, Pasadena, CA 91125, USA}%
\email{skefeli@caltech.edu}%
\author[0009-0003-5432-7180]{J.~M.~Kovac}%
\affiliation{Center for Astrophysics, Harvard \& Smithsonian, Cambridge, MA 01238, USA}%
\affiliation{Department of Physics, Harvard University, Cambridge, MA 02138, USA}%
\email{jmkovac@cfa.harvard.edu}%
\author{C.~L.~Kuo}%
\affiliation{Department of Physics, Stanford University, Stanford, CA 94305, USA}%
\email{clkuo@stanford.edu}%
\author[0000-0002-4540-1495]{K.~Lasko}%
\affiliation{School of Physics and Astronomy, University of Minnesota, Minneapolis, MN 55455, USA}%
\affiliation{Minnesota Institute for Astrophysics, University of Minnesota, Minneapolis, MN 55455, USA}%
\email{lasko062@umn.edu}%
\author[0000-0002-6445-2407]{K.~Lau}%
\affiliation{Department of Physics, California Institute of Technology, Pasadena, CA 91125, USA}%
\email{kennylau@caltech.edu}%
\author{M.~Lautzenhiser}%
\affiliation{Department of Physics, University of Cincinnati, Cincinnati, OH 45221, USA}%
\email{lautzemr@mail.uc.edu}%
\author[0000-0001-5677-5188]{T.~Liu}%
\affiliation{Department of Physics, Stanford University, Stanford, CA 94305, USA}%
\email{tongtianliu@stanford.edu}%
\author[0000-0002-1414-7236]{S.~C.~Mackey}%
\affiliation{Kavli Institute for Cosmological Physics, University of Chicago, Chicago, IL 60637, USA}%
\affiliation{Department of Physics, University of Chicago, Chicago, IL 60637, USA}%
\email{scmackey@uchicago.edu}%
\author{N.~Maher}%
\affiliation{School of Physics and Astronomy, University of Minnesota, Minneapolis, MN 55455, USA}%
\email{nojmaher@gmail.com}%
\author{K.~G.~Megerian}%
\affiliation{Jet Propulsion Laboratory, California Institute of Technology, Pasadena, CA 91109, USA}%
\email{krikor.g.megerian@jpl.nasa.gov}%
\author{L.~Minutolo}%
\affiliation{Department of Physics, California Institute of Technology, Pasadena, CA 91125, USA}%
\email{minutolo@caltech.edu}%
\author[0000-0002-4242-3015]{L.~Moncelsi}%
\affiliation{Department of Physics, California Institute of Technology, Pasadena, CA 91125, USA}%
\email{moncelsi@caltech.edu}%
\author{Y.~Nakato}%
\affiliation{Department of Physics, Stanford University, Stanford, CA 94305, USA}%
\email{yukanaka@stanford.edu}%
\author{H.~T.~Nguyen}%
\affiliation{Department of Physics, California Institute of Technology, Pasadena, CA 91125, USA}%
\affiliation{Jet Propulsion Laboratory, California Institute of Technology, Pasadena, CA 91109, USA}%
\email{hien.t.nguyen@jpl.nasa.gov}%
\author{R.~O’Brient}%
\affiliation{Department of Physics, California Institute of Technology, Pasadena, CA 91125, USA}%
\affiliation{Jet Propulsion Laboratory, California Institute of Technology, Pasadena, CA 91109, USA}%
\email{rogero@caltech.edu}%
\author{S.~N.~Paine}%
\affiliation{Center for Astrophysics, Harvard \& Smithsonian, Cambridge, MA 01238, USA}%
\email{spaine@cfa.harvard.edu}%
\author{A.~Patel}%
\affiliation{Department of Physics, California Institute of Technology, Pasadena, CA 91125, USA}%
\email{aapatel@caltech.edu}%
\author[0000-0002-4436-4215]{M.~A.~Petroff}%
\affiliation{Center for Astrophysics, Harvard \& Smithsonian, Cambridge, MA 01238, USA}%
\email{mpetroff@cfa.harvard.edu}%
\author[0000-0002-7822-6179]{A.~R.~Polish}%
\affiliation{Center for Astrophysics, Harvard \& Smithsonian, Cambridge, MA 01238, USA}%
\affiliation{Department of Physics, Harvard University, Cambridge, MA 02138, USA}%
\email{apolish@g.harvard.edu}%
\author{T.~Prouve}%
\affiliation{Service des Basses Temp\'eratures, Commissariat \`a l'\'Energie Atomique, 38054 Grenoble, France}%
\email{Thomas.PROUVE@cea.fr}%
\author[0000-0003-3983-6668]{C.~Pryke}%
\affiliation{School of Physics and Astronomy, University of Minnesota, Minneapolis, MN 55455, USA}%
\email{pryke@symmetryone.net}%
\author{C.~D.~Reintsema}%
\affiliation{National Institute of Standards and Technology, Boulder, CO 80305, USA}%
\email{strichte@gmail.com}%
\author{S.~Richter}%
\affiliation{Center for Astrophysics, Harvard \& Smithsonian, Cambridge, MA 01238, USA}%
\email{}%
\author{T.~Romand}%
\affiliation{Department of Physics, California Institute of Technology, Pasadena, CA 91125, USA}%
\email{tromand@caltech.edu}%
\author{M.~Salatino}%
\affiliation{Department of Physics, Stanford University, Stanford, CA 94305, USA}%
\email{marias5@stanford.edu}%
\author{A.~Schillaci}%
\affiliation{Department of Physics, California Institute of Technology, Pasadena, CA 91125, USA}%
\email{aleschillaci78@gmail.com}%
\author{B.~Schmitt}%
\affiliation{Center for Astrophysics, Harvard \& Smithsonian, Cambridge, MA 01238, USA}%
\email{benjamin.schmitt@cfa.harvard.edu}%
\author{R.~Schwartz}%
\affiliation{School of Physics and Astronomy, University of Minnesota, Minneapolis, MN 55455, USA}%
\email{iceman@antarctic{-}adventures.de}%
\author{C.~D.~Sheehy}%
\affiliation{School of Physics and Astronomy, University of Minnesota, Minneapolis, MN 55455, USA}%
\email{chris.sheehy@descarteslabs.com}%
\author[0000-0001-7387-0881]{B.~Singari}%
\affiliation{School of Physics and Astronomy, University of Minnesota, Minneapolis, MN 55455, USA}%
\affiliation{Minnesota Institute for Astrophysics, University of Minnesota, Minneapolis, MN 55455, USA}%
\email{singa044@umn.edu}%
\author{A.~Soliman}%
\affiliation{Department of Physics, California Institute of Technology, Pasadena, CA 91125, USA}%
\affiliation{Jet Propulsion Laboratory, California Institute of Technology, Pasadena, CA 91109, USA}%
\email{asoliman@caltech.edu}%
\author{T.~St.~Germaine}%
\affiliation{Center for Astrophysics, Harvard \& Smithsonian, Cambridge, MA 01238, USA}%
\email{tstgermaine7@gmail.com}%
\author[0000-0003-0260-605X]{A.~Steiger}%
\affiliation{Department of Physics, California Institute of Technology, Pasadena, CA 91125, USA}%
\email{asteiger@caltech.edu}%
\author{B.~Steinbach}%
\affiliation{Department of Physics, California Institute of Technology, Pasadena, CA 91125, USA}%
\email{bsteinba@caltech.edu}%
\author{R.~Sudiwala}%
\affiliation{School of Physics and Astronomy, Cardiff University, Cardiff, CF24 3AA, United Kingdom}%
\email{R.Sudiwala@astro.cf.ac.uk}%
\author{G.~Teply}%
\affiliation{Department of Physics, California Institute of Technology, Pasadena, CA 91125, USA}%
\email{gteply@ucsd.edu}%
\author{K.~L.~Thompson}%
\affiliation{Department of Physics, Stanford University, Stanford, CA 94305, USA}%
\affiliation{Kavli Institute for Particle Astrophysics and Cosmology, Stanford University, Stanford, CA 94305, USA}%
\email{Keith.L.Thompson@stanford.edu}%
\author[0000-0002-1851-3918]{C.~Tucker}%
\affiliation{School of Physics and Astronomy, Cardiff University, Cardiff, CF24 3AA, United Kingdom}%
\email{carole.tucker@astro.cf.ac.uk}%
\author{A.~D.~Turner}%
\affiliation{Jet Propulsion Laboratory, California Institute of Technology, Pasadena, CA 91109, USA}%
\email{anthony.d.turner@jpl.nasa.gov}%
\author{A.~G.~Vieregg}%
\affiliation{Kavli Institute for Cosmological Physics, University of Chicago, Chicago, IL 60637, USA}%
\affiliation{Department of Physics, University of Chicago, Chicago, IL 60637, USA}%
\email{avieregg@kicp.uchicago.edu}%
\author[0000-0002-8232-7343]{A.~Wandui}%
\affiliation{Department of Physics, California Institute of Technology, Pasadena, CA 91125, USA}%
\email{awandui@caltech.edu}%
\author{A.~C.~Weber}%
\affiliation{Jet Propulsion Laboratory, California Institute of Technology, Pasadena, CA 91109, USA}%
\email{Alexis.C.Weber.Dr@jpl.nasa.gov}%
\author[0000-0002-6452-4693]{J.~Willmert}%
\affiliation{School of Physics and Astronomy, University of Minnesota, Minneapolis, MN 55455, USA}%
\email{justin@willmert.me}%
\author{C.~L.~Wong}%
\affiliation{Center for Astrophysics, Harvard \& Smithsonian, Cambridge, MA 01238, USA}%
\affiliation{Department of Physics, Harvard University, Cambridge, MA 02138, USA}%
\email{chinlinnet@gmail.com}%
\author[0000-0001-5411-6920]{W.~L.~K.~Wu}%
\affiliation{Kavli Institute for Particle Astrophysics and Cosmology, Stanford University, Stanford, CA 94305, USA}%
\affiliation{SLAC National Accelerator Laboratory, Menlo Park, CA 94025, USA}%
\email{wlwu@slac.stanford.edu}%
\author{H.~Yang}%
\affiliation{Department of Physics, Stanford University, Stanford, CA 94305, USA}%
\email{ericy4712@gmail.com}%
\author[0000-0002-8542-232X]{C.~Yu}%
\affiliation{Kavli Institute for Cosmological Physics, University of Chicago, Chicago, IL 60637, USA}%
\affiliation{High-Energy Physics Division, Argonne National Laboratory, Lemont, IL, 60439, USA}%
\email{cyndiayu@uchicago.edu}%
\author[0000-0001-6924-9072]{L.~Zheng}%
\affiliation{Center for Astrophysics, Harvard \& Smithsonian, Cambridge, MA 01238, USA}%
\email{lingzhen.zeng@cfa.harvard.edu}%
\author[0000-0001-8288-5823]{C.~Zhang}%
\affiliation{Kavli Institute for Particle Astrophysics and Cosmology, Stanford University, Stanford, CA 94305, USA}%
\email{chzzhang@stanford.edu}%
\author{S.~Zhang}%
\affiliation{Department of Physics, California Institute of Technology, Pasadena, CA 91125, USA}%
\email{szhang5@caltech.edu}

\begin{abstract}
The BICEP series of experiments has been observing CMB polarisation from the South Pole for over 20 years, with the goal of constraining inflationary gravitational waves.
The upcoming data release, using data taken through 2024, is forecasted to constrain the tensor-to-scalar ratio $r$ at the level of $\sigma(r) \sim 0.005$ (including delensing), with the longer-term goal of reaching $\sigma(r) \sim 0.001$ by 2034. 
As the survey sensitivity increases, it is crucial to control instrumental systematics to unprecedented levels, and our goal is to limit dominant sources of systematics to 20\% of $\sigma(r)$ or lower. 
Achieving this requires careful instrumental characterisation and dedicated end-to-end studies to evaluate the impact of each systematic on cosmological parameters. 
We first present the BICEP calibration program, which characterises the optical, spectral, and polarisation response of the receivers. 
We then describe the analysis strategies implemented to identify and mitigate systematic contamination, and we detail simulations used to evaluate the impact of residual effects.
We focus in particular on beam systematics, the dominant source of systematics for BICEP receivers. 
Finally, we report preliminary estimates of the expected level of systematic contamination for upcoming BICEP results, and we discuss approaches to evaluate and mitigate instrumental systematics for future surveys.
\end{abstract}


\section{The BICEP program}
The BICEP program has been continuously observing from the South Pole for two decades to characterise CMB polarisation. 
The current experimental configuration consists of five small-aperture receivers spanning six frequency bands from 30 to 270 GHz~\cite{B3,BA}.
This multi-frequency approach is essential for the joint estimation of the primordial CMB signal and the galactic foreground signal.
Using data acquired through the 2018 season alongside WMAP and \textit{Planck} observations, the BICEP collaboration established the most stringent constraints on primordial gravitational waves to date, reaching a sensitivity of $\sigma(r) = 0.009$ and setting an upper limit on the tensor-to-scalar ratio of $r < 0.036$ at 95\% confidence~\cite{BK18}.
As raw instrumental sensitivity increases with receiver upgrades and extended integration time, we will additionally leverage high-resolution measurements from the South Pole Telescope (SPT) to perform delensing— removal of gravitational lensing induced B-modes~\cite{BKSPT}.
The forthcoming BK24 + SPT dataset, using BICEP observations through 2024, is forecasted to reach $\sigma(r) \sim 0.005$.
Looking toward the next decade, the joint South Pole Observatory (SPO) program, combining BICEP and SPT observations, is projected to achieve $\sigma(r)\lesssim 0.001$ by 2034~\cite{HEPAP2024}.

As sensitivity increases, control of instrumental systematics becomes crucial to ensure unbiased results. 
Previous studies estimated dominant systematics at the level of $r \sim 0.001 \mbox{--} 0.002$~\cite{BK18,BKIII,BKXI}, but these estimates must be updated for new receivers and more sensitive datasets. 
Moreover, it is essential to account for the multi-frequency nature of the data and propagate sources of contamination through a multicomponent analysis to obtain reliable estimates for each effect in each band. 
We rely on a full-loop approach to identify and mitigate systematic effects, combining intensive calibration campaigns, simulations informed by per-detector measurements, and multicomponent analysis to set requirements based on bias on \textit{r}.

\section{Experimental approach to systematics}
\subsection{Design choices}

The BICEP experimental approach incorporates a number of built-in hardware~\cite{B3,BKIV} and analysis~\cite{BKV} strategies designed to mitigate instrumental systematics.
Receivers have an on-axis optical design with an under-illuminated aperture to minimize diffraction, while the small aperture size facilitates comprehensive ground-based calibration. 
The design features a co-moving absorptive forebaffle paired with a reflective ground shield to minimise stray light contamination and ground pickup.
Furthermore, we employ a repetitive, small-patch scanning strategy, which enables targeted jackknife consistency tests that are essential in identifying systematics.

\subsection{Calibration campaigns}

To obtain the high-quality, high-SNR measurements necessary for the required level of systematics control, we conduct intensive on-site calibration campaigns. 
A cornerstone of this strategy is an annual 6–8 week far-field beam mapping (FFBM) campaign to characterise the main-beam temperature response of every detector in continuously upgraded BICEP receiver~\cite{SPIE2020,SPIE2024}.
These measurements are integral in controlling temperature-to-polarisation leakage (section~\ref{sec:tpleakage}).
All calibrations are performed in situ under conditions nearly identical to CMB observations, including measurements of near field beam response, bandpasses, polarization angles and efficiencies~\cite{BKXVIII}, as well as temperature and polarised beam maps extending far beyond the main beam~\cite{SPIE2024}.

\section{Treatment of temperature-to-polarisation leakage}
\label{sec:tpleakage}
BICEP receivers rely on pair-differencing between orthogonal, linearly polarised detectors to reject unpolarised atmospheric contamination and reconstruct the polarised sky signal. 
Differences between paired detectors can lead to incomplete cancellation of the temperature signal, resulting in temperature-to-polarisation ($T \rightarrow P$) leakage. 
Among the possible sources, differential main beam response has long been identified as the dominant effect, requiring careful mitigation~\cite{BKIII}.

\subsection{Deprojection}
To mitigate $T \rightarrow P$ leakage, we employ the deprojection technique.
This method fits for the amplitude of leakage templates arising from differential beam response, and removes the contaminated modes at the map-making stage~\cite{BKIII}.
For past data releases, we deprojected the lowest-order, physically motivated modes of the differential beam: gain mismatch, differential pointing, and differential ellipticity.
While other effects have been identified in beam maps, they previously did not cause significant contamination.
For BK24, jackknife tests on the deepest map (BICEP3 9-year) have revealed residual EE contamination due to a combination of illumination ellipticity and beam truncation at the aperture stop ("wide quadrupole" effect). 
We similarly identified contamination due to readout crosstalk, which is of electrical origin, but can be described in the same framework.
This calls for additional mitigation to ensure that residual systematic biases remain below the target sensitivity.

\subsection{Beam simulations}
\begin{figure}[ht!]
    \centering
    \includegraphics[width = 0.6\textwidth]{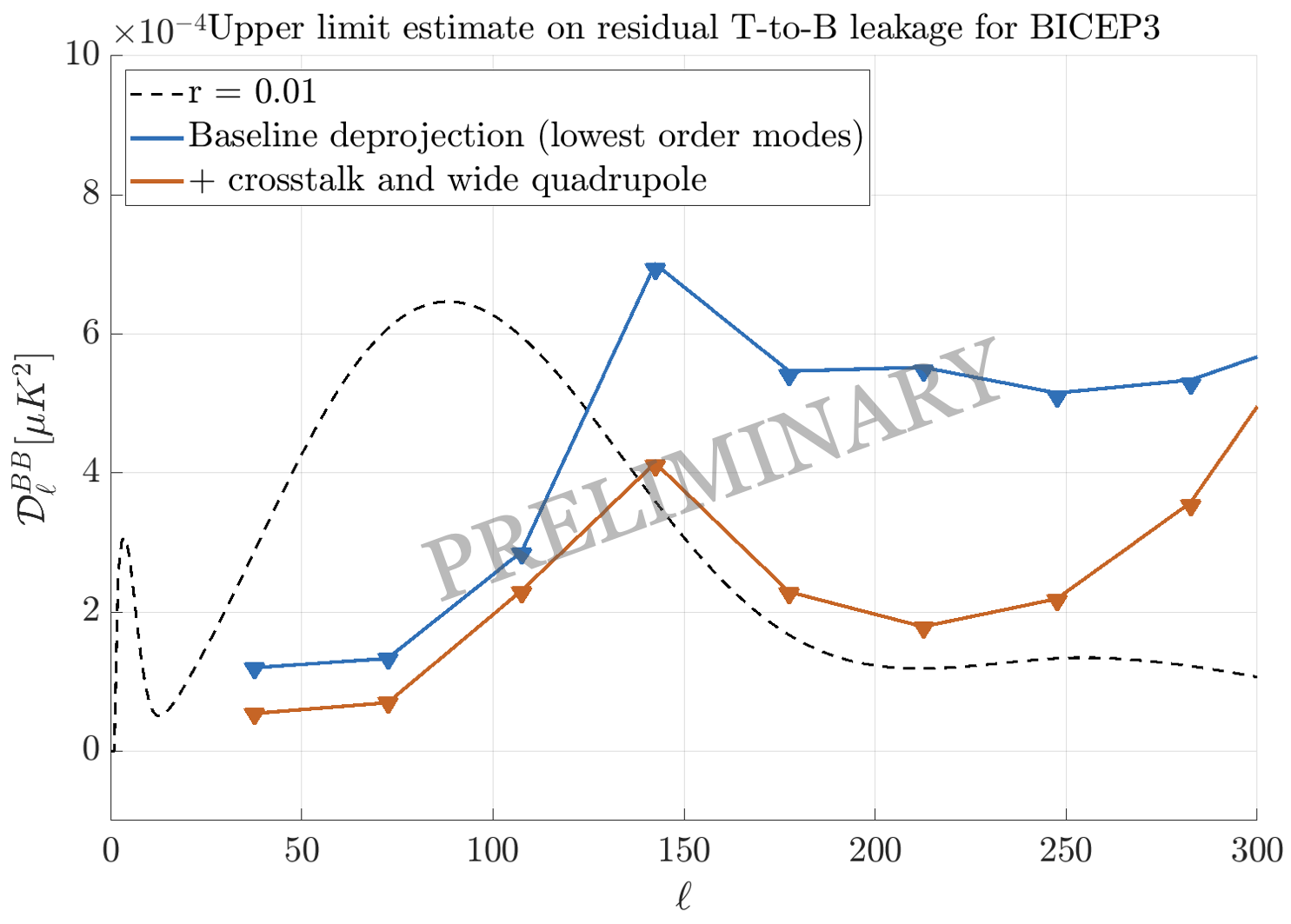}
    \caption{Upper limit on BB temperature-to-polarisation leakage residual for the 9-year BICEP3 map (95 GHz)}
    \label{fig:B3}
\end{figure}

After deprojecting leading-order $T \rightarrow P$ leakage modes, we are left with residuals whose amplitude and impact on $r$ must be assessed.
We estimate the undeprojected residuals using specialised, high-fidelity simulations in which measured detector beam maps are convolved with a temperature-only sky map, naturally incorporating beam effects present in the real data.
The simulated data are then processed through the same analysis pipeline as the real data --- including cuts, filtering and deprojection~\cite{BKIII,BKXI}.
The resulting polarisation signal provides an estimate of the $T \rightarrow P$ leakage in real data. 
This estimate should be interpreted as an upper limit, since it may be biased high by measurement systematics in the beam mapping data.
Figure~\ref{fig:B3} shows preliminary results for the BICEP3 9-year map. 
We find that additional deprojection of physically motivated modes (crosstalk and wide quadrupole) effectively reduces the residuals, and we continue to improve this procedure for the final data release.

\section{Multicomponent simulations and analysis}

\subsection{Framework}
Once we have estimates of residual $T \rightarrow P$ leakage and other instrumental systematics, we propagate them through a multicomponent analysis framework. 
We estimate best-fit parameters through a maximum likelihood search, evaluating the data against a model including the primordial CMB, lensing signal, and galactic dust and synchrotron~\cite{BK18}.
To assess the impact of systematics, we start from systematics-free simulations and add systematic contamination to derive the bias on $r$ on a realisation-to-realisation basis. 
In addition to the BICEP end-to-end simulation pipeline~\cite{BKII}, we have developed a bandpower-based framework using expectation values and Gaussian noise, based on the achieved or forecasted bandpower covariance matrix~\cite{Eiben}.
This enables rapid turnaround, allowing iteration on requirements for ongoing analysis, and straightforward scaling to future data sets.

\subsection{Preliminary results}
We use this framework to derive the bias on $r$ from various sources of contamination, to ensure that no effect produces a bias larger than 20\% of $\sigma(r)$, i.e., $\Delta(r) < 0.001$ for BK24.
In particular, we set constraints on the level of residual $T \rightarrow P$ leakage by observing band.
We show that most bands largely meet requirements, and that BICEP3 requires additional mitigation of readout crosstalk and wide quadrupole. Additionally, we establish upper limits on bias on $r$ induced by bandcenter shifts, by simulating cumulative $\pm 2\%$ shifts, a conservative upper limit on bandpass measurement uncertainty.
We show that this level of control yields $\Delta(r) = 0.0009$, which meets our requirement.
Other evaluated systematics (polarisation angles and cross-polar response) are well below the threshold, and we are currently assessing additional effects.

This multicomponent framework provides insight into how individual bands contribute to the systematic budget based on their sensitivity, sky coverage, and observing frequency. 
An example is given in Table~\ref{tab:bandpass}, showing how the same bandcenter shift can yield to orders-of-magnitude differences in bias on $r$ depending on the band. 
This type of result guides future calibration campaigns by identifying which systematics require tighter control in each band.

\begin{deluxetable}{lcccccccc}[ht!]
\tablecaption{Per-band bias on $r$ from a +2\% bandcenter shift. \label{tab:bandpass}}
\tablewidth{\columnwidth}      
\tablehead{
\colhead{\textbf{Band [GHz]}} & 
\colhead{\textbf{30}} & 
\colhead{\textbf{40}} & 
\colhead{\textbf{\shortstack{95\\(Keck)}}} & 
\colhead{\textbf{\shortstack{95\\(BICEP)}}} & 
\colhead{\textbf{\shortstack{150\\(Keck)}}} & 
\colhead{\textbf{\shortstack{150\\(BICEP)}}} & 
\colhead{\textbf{220}} & 
\colhead{\textbf{270}}
}
\startdata
$\Delta(r) \times 10^{-4}$ & 0.04 & 0.02 & $-0.3$ & $-3.7$ & 0.3 & 3.5 & 2.7 & $-0.6$ \\
\enddata
\end{deluxetable}

\section{Conclusion}

As the sensitivity of CMB experiments improves, controlling instrumental systematics becomes increasingly critical for obtaining unbiased constraints on inflation.
In this work, we have presented the BICEP calibration and analysis framework for characterising, mitigating, and propagating instrumental effects to cosmological parameters.
We demonstrate that dominant systematic contributions can be controlled to $\Delta(r) \lesssim 0.001$ for the upcoming BK24 data set. 
This relies on a full-loop approach that integrates individual detector measurements from intensive calibration campaigns into high-fidelity simulations and systematic evaluations. 
The multicomponent framework provides a flexible and scalable approach for evaluating systematics across frequency bands and experimental configurations. These tools will be essential for future analyses targeting even higher sensitivity in the coming decade.

\begin{acknowledgments}
The BICEP/Keck projects have been made possible
through a series of grants from the NSF, including 2220444-2220448, 2216223, 1836010, \& 1726917. The development of antenna-coupled detector technology was supported by the JPL Research and Technology Development Fund and by NASA Grants 06-ARPA206-0040, 10-SAT10-0017, 12-SAT12-0031, 14-SAT14-0009, \& 16-SAT-16-0002. The development and testing of focal planes was supported by the Gordon and Betty Moore Foundation at Caltech. Readout electronics were supported by a Canada Foundation for Innovation grant to UBC. Support for quasi-optical filtering was provided by UK STFC grant ST/N000706/1. The computations in this paper were run on the Cannon cluster supported by the FAS Research Computing Group at Harvard University. This work was supported by the DOE Office for Science, award number DE-AC02-05CH11231, and LBNL Laboratory-Directed Research \& Development funds.
\end{acknowledgments}

\bibliography{references}{}
\bibliographystyle{unsrtnat}

\end{document}